\documentclass[12pt,a4paper,twoside]{article}

\usepackage{authblk}

\usepackage[T1]{fontenc}

\usepackage{mathrsfs}
\usepackage[english]{babel} 
\usepackage{bbm}
\usepackage{bm}
\usepackage{amsmath,amsfonts,amscd,amssymb,latexsym}
\usepackage{wrapfig}

\usepackage{graphics}
\graphicspath{{/}}
\DeclareGraphicsExtensions{.pdf,.png,.jpg}

\usepackage{rotating}
\usepackage{cite}
\usepackage{geometry} %
\def\kappadp{\kappa_{D \rightarrow p}}

\usepackage[hang, small,labelfont=bf,up,textfont=up]{caption} 
\usepackage{booktabs} 
\usepackage{float} 
\usepackage{hyperref} 

\usepackage{lettrine} 
\usepackage{paralist} 

\usepackage{abstract} 

\usepackage{titlesec} 
\titleformat{\section}[block]{\large\scshape}{\thesection.}{1em}{} 
\titleformat{\subsection}[block]{\large}{\thesubsection.}{1em}{} 

\title{\fontsize{16pt}{12pt}\selectfont\textbf{On the proton-deuteron backward scattering}} 

\author[ ]{S. Bondarenko\thanks{Corresponding Author: bondarenko@jinr.ru}}
\author[ ]{S. Yurev}
\affil[ ]{\small{Bogoliubov Laboratory of Theoretical Physics, Joint Institute for Nuclear Research,
141980 Dubna, Russia}}

\date{}
\begin{document}

\begin{center}
\fontsize{16pt}{12pt}\selectfont\textbf{On the proton-deuteron backward scattering}
\end{center}

\begin{center}
S. Bondarenko*, S. Yurev
\end{center}

\begin{center}
\small{Bogoliubov Laboratory of Theoretical Physics,

Joint Institute for Nuclear Research, 141980 Dubna, Russia

*bondarenko@jinr.ru}
\end{center}

\begin{abstract}

The article is devoted to the relativistic study of elastic $pD$ backward scattering based on the one-nucleon exchange diagram.
Calculations were performed using 
relativistic 
deuteron wave functions obtained by solving the Bethe-Salpeter 
equation in the Minkowski space with relativistic separable potentials.
The unpolarized differential cross section as well as some polarization observables of the reaction for initial proton momentum up to 7.3 GeV were calculated.
The obtained results are compared with the calculations of other authors.

\end{abstract}
	
\section{Introduction}
The study of the elastic proton-deuteron scattering reaction is one of the 
important problems in nuclear physics. 
To describe the reaction, one needs to consider the unbound three-nucleon system.
Technically, this task is more difficult than the consideration of three-nucleon bound states since the
scattering energy of the $pD$-system is unlimited.
It is especially interesting to investigate  backward scattering.
In this case, in the one-nucleon exchange approximation the nonrelativistic formulae for the unpolarized cross section is proportional to the sum
of the radial parts of the $S$ and $D$ partial-wave states of the deuteron wave function squared.
So it was  hoped to extract the distribution of  nucleons in the deuteron directly from  experimental data.
However, all models for nucleon-nucleon (NN) potentials  failed to describe the experimental data.

In the nonrelativistic case, the modern theoretical study of $pD$ scattering is carried out in most cases by solving the system of Faddeev integral equations. In the relativistic case the study of  elastic $pD$ backward scattering
in the framework of the Faddeev approach is not a popular method due to the limited number of relevant relativistic generalizations and  technical difficulties associated with specific numerical calculations. One of the promising approaches in this direction may be the Bethe-Salpeter-Faddeev approach within the separable
kernel of the NN interaction proposed in the article~\cite{Rupp:1991th} and developed in~\cite{Bondarenko:2020, Bondarenko:2021} as applied to three-nucleon nuclei.

One of the most common approaches to the study of $pD$ scattering in the relativistic case is the analysis of the Feynman interaction diagram based on one-nucleon exchange, as well as corrections of the higher-order diagrams, allowing for the $\pi$-meson exchange~\cite{KD98}, $\Delta $ isobars~\cite{LA2020} and so on.
The scattering amplitude includes the deuteron wave function, which can be found, for example, by solving the Bethe-Salpeter (BS) equation. The elastic scattering cross section is proportional to the 4th power of the 
deuteron wave function and therefore the result is very sensitive to the choice of this function. One can limit oneself to the deuteron $S$ and $D$ partial-wave states, or even take into account the $P$-states. This method is developed in some works, for example,~\cite{KD98,LA2020}.

There are a lot of experimental data on proton-deuteron scattering for the unpolarized cross section as well as for
some polarization characteristics of the reaction. The experimental results are also obtained at high energies, 
which requires an appropriate theoretical relativistic consideration of the problem.

The study of polarization characteristics makes it possible to answer the question of the influence of spin and orbital momenta on nucleon-nucleon interactions. Among the important polarization quantities, one can single out the tensor analyzing power $T_{20}$ and the  polarization  transfer $\kappadp$ from the initial deuteron to the final proton, for which there is a large amount of experimental data. Also, important polarization characteristics are 
the vector-vector polarization transfer (VVPT) coefficient from the initial proton to the final proton $H_{N,0 \rightarrow N,0}$, 
the vector-vector polarization transfer coefficient from the initial deuteron to the final deuteron $H_{0,N \rightarrow 0,N}$ and
the tensor-tensor polarization transfer (TTPT) coefficient from the initial deuteron to the final deuteron $H_{0,NN \rightarrow 0,SS}$ for which there are no experimental data yet.

In the paper the influence of some relativistic kernels of the NN interaction is considered 
on the unpolarized cross section and polarization observables -- relativistic Graz-II~\cite{GRAZ} and
MY6\cite{MY6} in the one-nucleon exchange approximation. The influence of the zeroth component 
of the relative momentum in the BS amplitude is also investigated.

The paper is organized as follows: in Sec. 2, the formalism is shortly written.
In Sec. 3, the results of calculations and discussion are presented, and
in Sec. 4, the summary is given.

\section{Formalism}

In the one-nucleon exchange approximation the covariant amplitude of the elastic $pD$ backward scattering has the following form:  
\begin{eqnarray}
\mathcal{M}  = 
\overline{u}(p_{f},s_{f}) \Gamma_{M_i} (D_{i}, q_{i}) {\tilde S}_2 \overline{\Gamma}_{M_f}(D_{f}, q_{f}) u(p_{i},s_{i}),
\label{amp}
\end{eqnarray}
where the initial (final) protons with 4-momenta
$p_{i}$ ($p_{f}$) and polarization $s_{i}$ ($s_{f}$) 
are described by the Dirac bispinor functions and 
the initial (final) deuterons with total
$P_{i}$ ($P_{f}$) and relative $q_{i} (q_{f})$  4-momenta by the vertex functions $\Gamma (\overline{\Gamma})$; ${\tilde S}_2 = 1/(D\cdot\gamma - q\cdot\gamma + m)$, ($m$ is the nucleon mass) is the intermediate-nucleon modified propagator where $\gamma$ are the Dirac
matrices. The relative 4-momenta are: $q_i = p_f - D_i/2$ and  $q_f = p_i - D_f/2$. 
The reaction is considered in the laboratory system where $D_i=(M_d,0)$
and $p_f = (E_f,{\bm p_f})$ ($E_f = \sqrt{m^2+{\bm p_f}^2}$, $M_d$ is deuteron mass).

The initial and final proton momenta are related as follows:
\begin{eqnarray}
|{\bm p_f}| = \frac{M_d^2 - m^2}{s}|{\bm p_i}|,
\label{out_in}
\end{eqnarray}
where $s = (D_i+p_i)^2 = M_d^2 + m^2 + 2 M_d E_i$ ($E_i = \sqrt{m^2+{\bm p_i}^2}$) is the total momentum squared. 
From Eq.~(\ref{out_in}) one can obtain that if $|{\bm p_i}| \to \infty$
then $|{\bm p_f}| \to (M_d^2 - m^2)/(2 M_d) \approx 0.702$ GeV/c is the kinematic limit.

To calculate observables (cross section and polarization characteristics), one needs to square the amplitude~(\ref{amp})
and change the $u{\overline{u}}$ and $\Gamma \overline{\Gamma}$ quantities to the density matrices
which correspond to the polarization characteristics under consideration. This technique is described in detail in~\cite{KD98,LL96}. In this paper the expressions given in~\cite{KD98} are used for calculation.
Moreover, to  clearly see the impact of different NN interactions kernels,
the relativistic effects caused by Lorentz boosts are omitted in the paper.

It should be noted that, the arguments in the radial parts of the deuteron BS vertex function should be calculated 
in their rest frame systems, and therefore
$|{\bm q_i}| = |{\bm q_f}| = |{\bm p_f}|$ and $q_{f0} = q_{i0} = {\tilde q}_0 = E_f - M_d/2$.
If a solution of the BS equation is obtained in the pseudo-Euclidean space~\cite{KT82,KD98}, the analytic continuation 
into the complex $q_0$ plane should be carried out. In this case, usually $q_0$ is set equal to zero.
In the paper, the separable kernel of the NN interaction is used to solve the BS equation in the Minkowski space, and the
exact dependence on  $q_0 = {\tilde q}_0$ can be taken into account.

\begin{figure}[!htb]
\center{
\begin{tabular}{cc}
    \includegraphics[width=0.45\linewidth,angle=0]{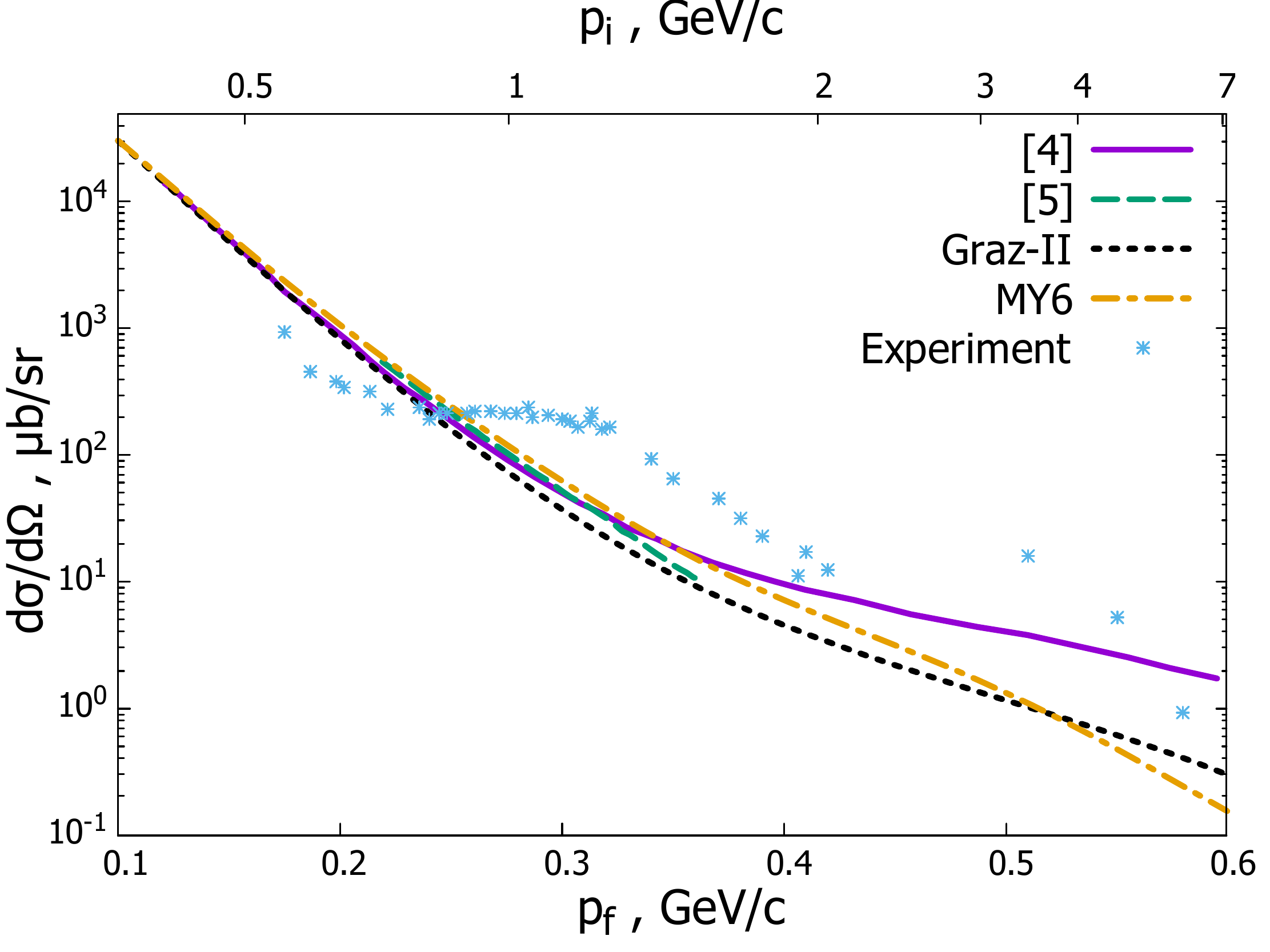} &
	\includegraphics[width=0.45\linewidth,angle=0]{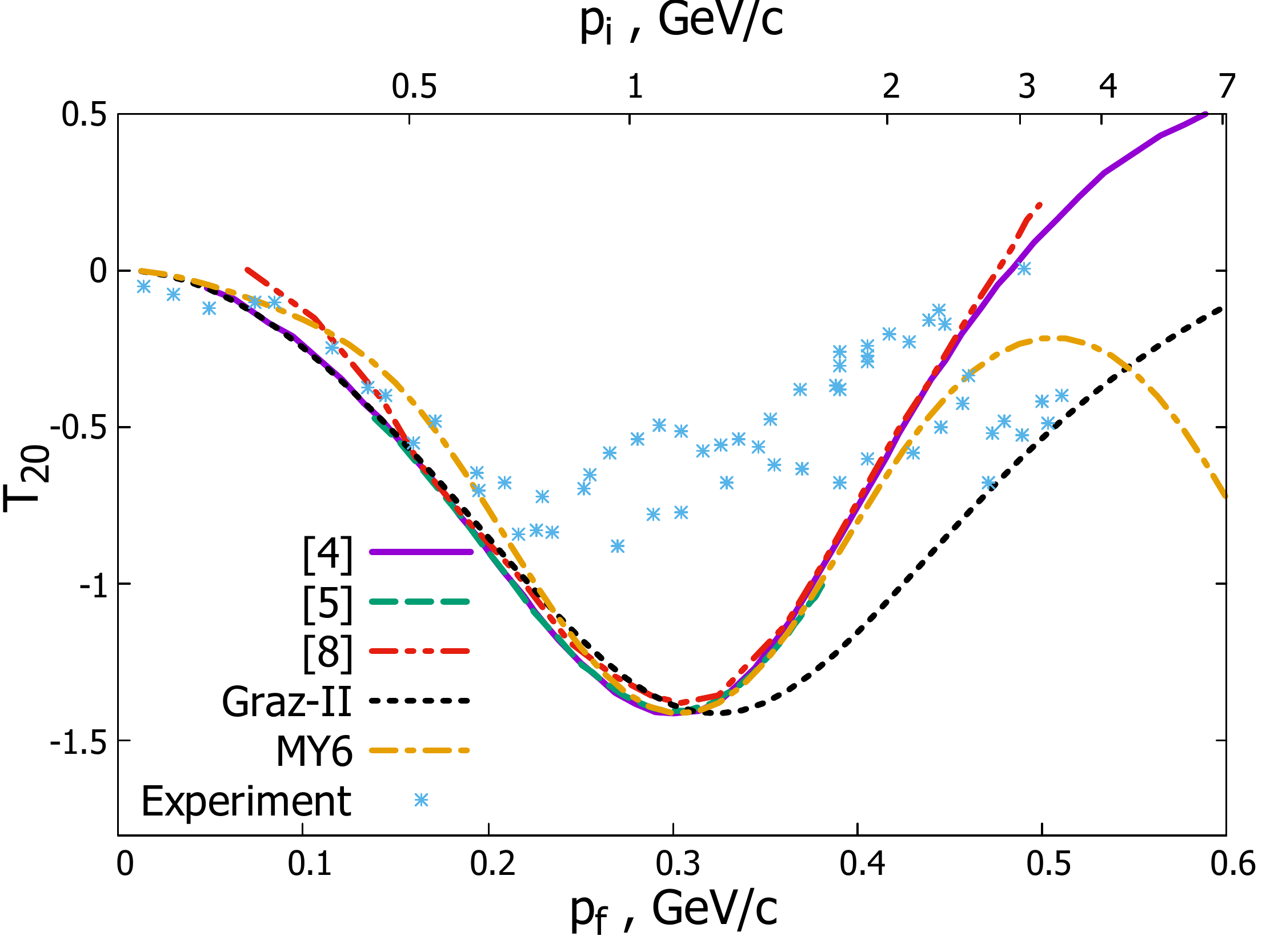} \\
	\includegraphics[width=0.45\linewidth,angle=0]{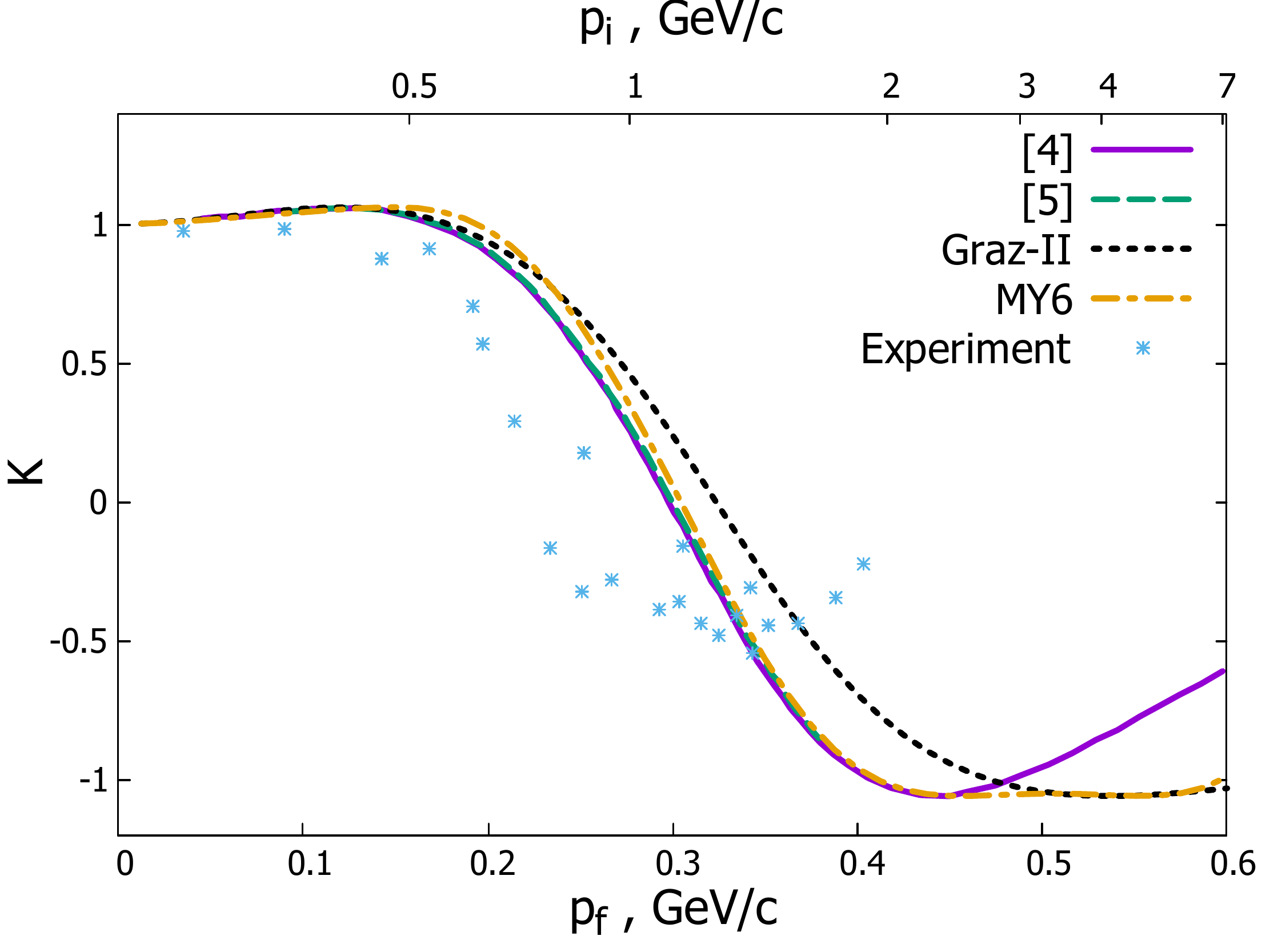} &
	\includegraphics[width=0.45\linewidth,angle=0]{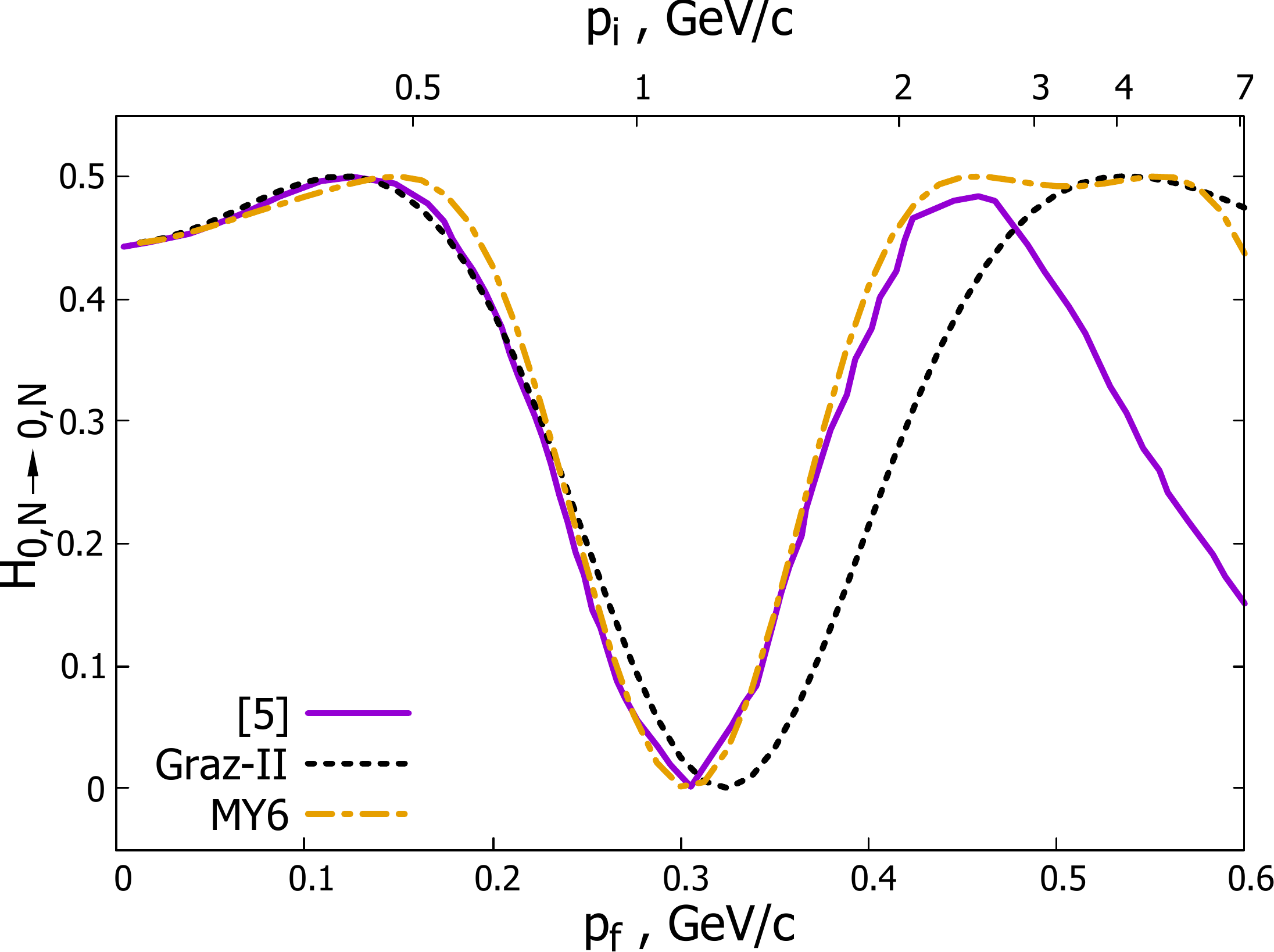}\\
	\includegraphics[width=0.45\linewidth,angle=0]{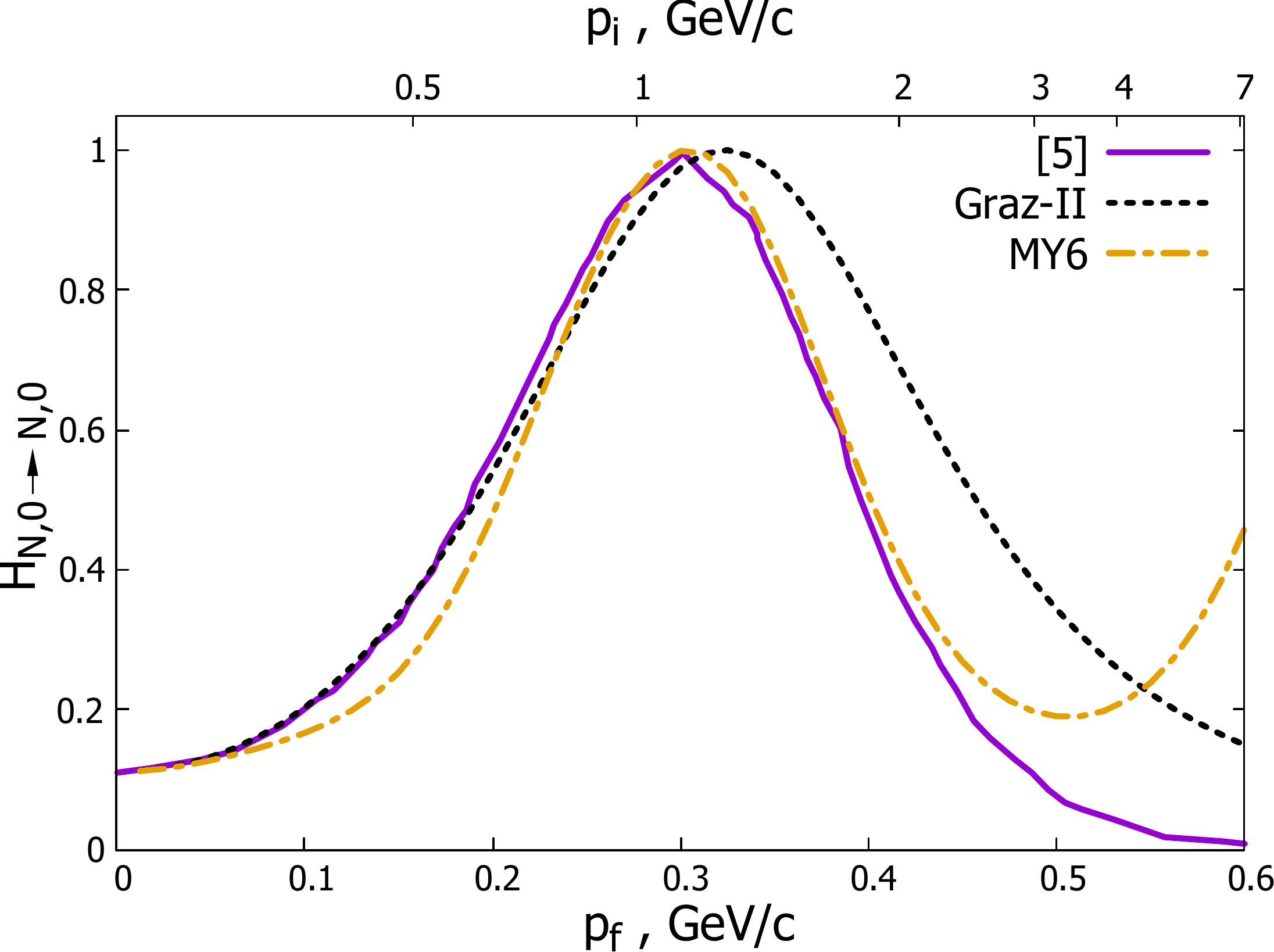}&
	\includegraphics[width=0.45\linewidth,angle=0]{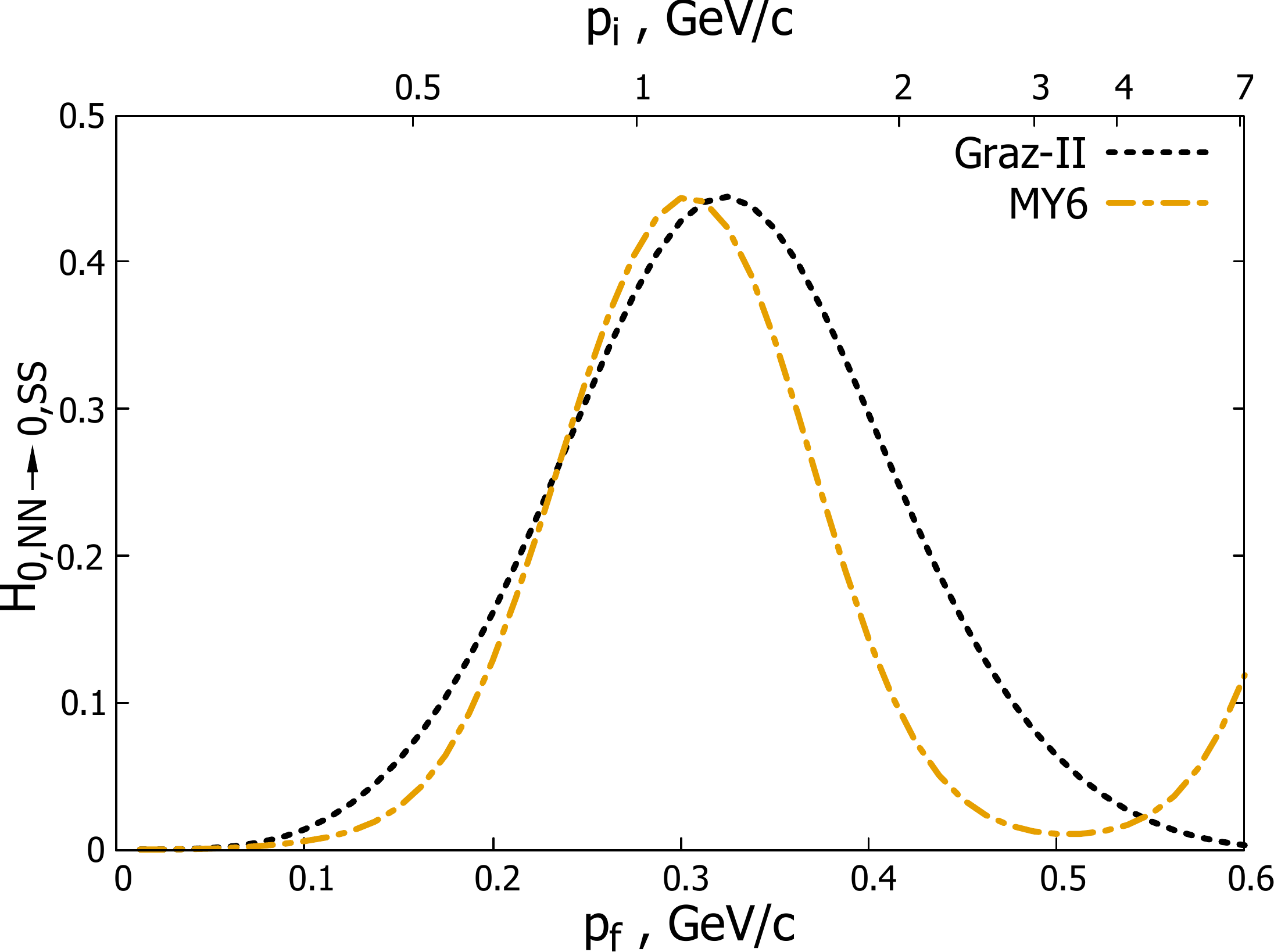}
\end{tabular}
}
\caption{Unpolarized differential cross section (left upper),
tensor analyzing power $T_{20}$ (right upper), polarization  transfer  $\kappadp$ (left middle),
VVPT coefficient from the initial to the final deuteron 
$H_{0,N \rightarrow 0,N}$ (right middle)
and the VVPT coefficient from the initial to the final proton 
$H_{N,0 \rightarrow N,0}$ (left bottom),
TTPT coefficient from the initial to the final deuteron $H_{0,NN \rightarrow 0,SS}$ (right bottom).
Calculations from~\cite{KD98} are shown by solid line, from~\cite{KT82} -- by dashed-dotted-dotted line, from~\cite{LA2020} -- by dashed line, 
from this article with potential Graz-II -- by dotted line and with MY6 potential -- by dashed-dotted line. 
Experimental data taken from~\cite{KD98} and references from it.
}
\label{fig1}
\end{figure}	
\begin{figure}[!htb]
\center{
\begin{tabular}{cc}    
	\includegraphics[width=0.45\linewidth,angle=0]{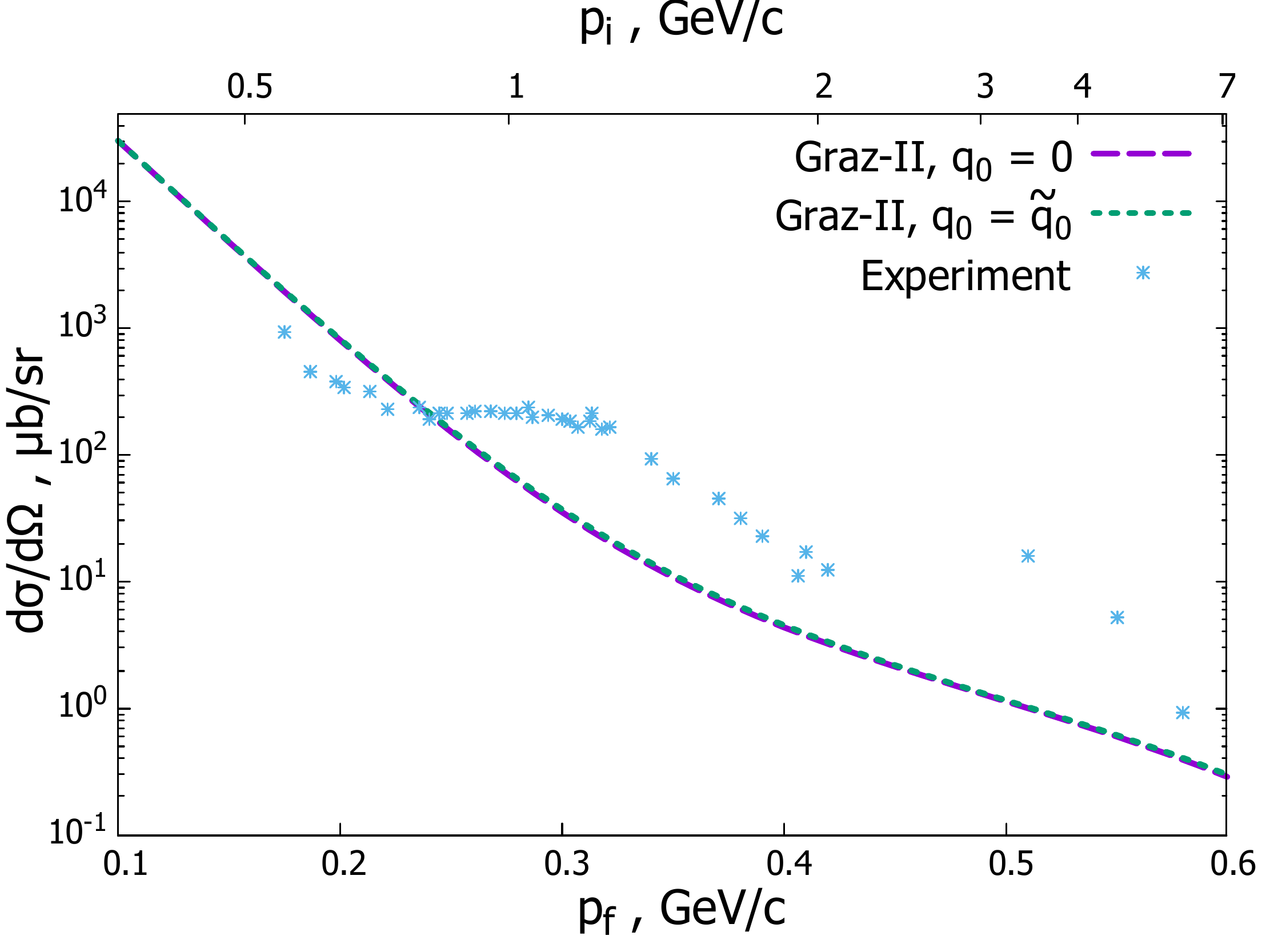}&
	\includegraphics[width=0.45\linewidth,angle=0]{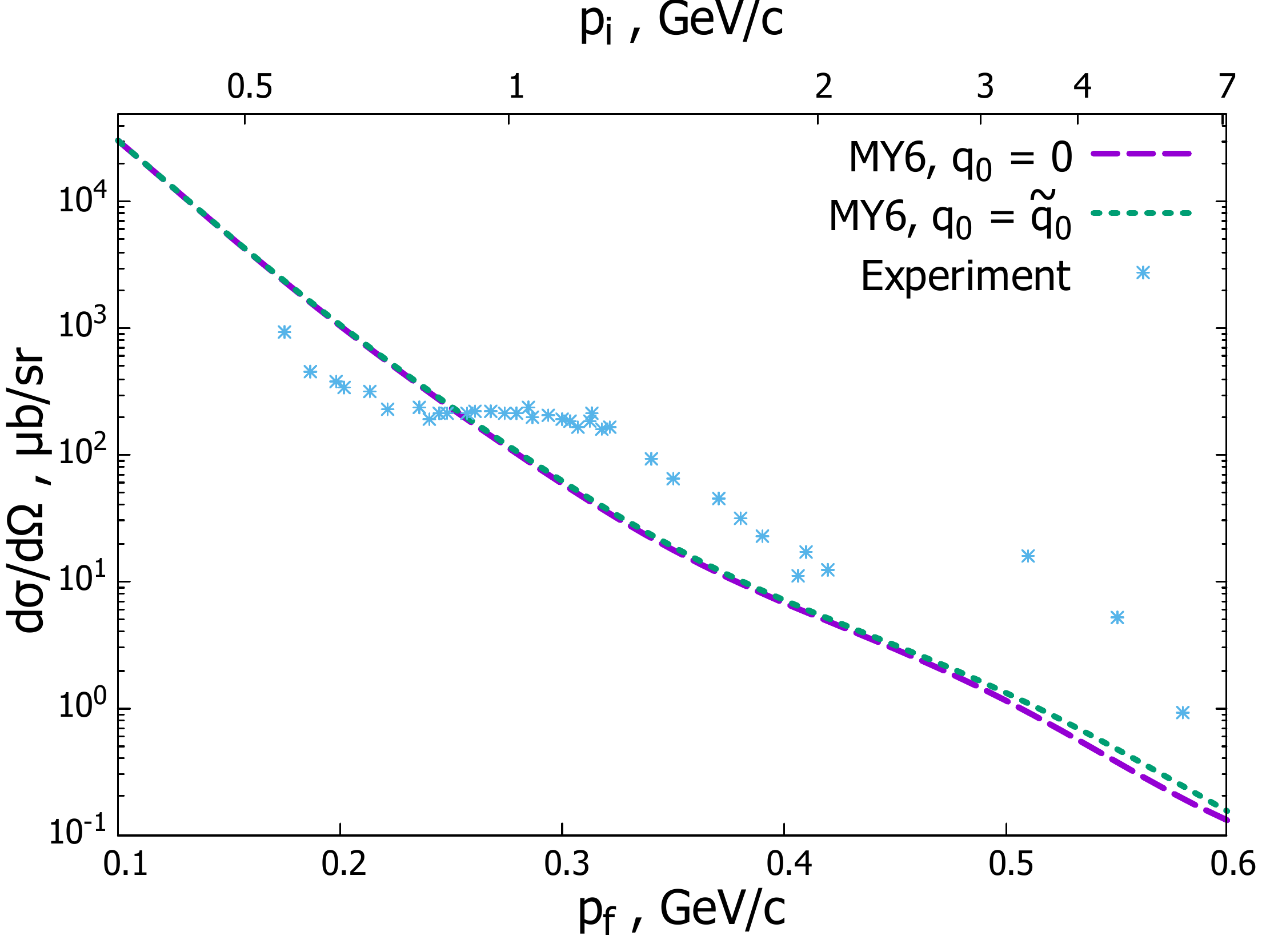}\\
	\includegraphics[width=0.45\linewidth,angle=0]{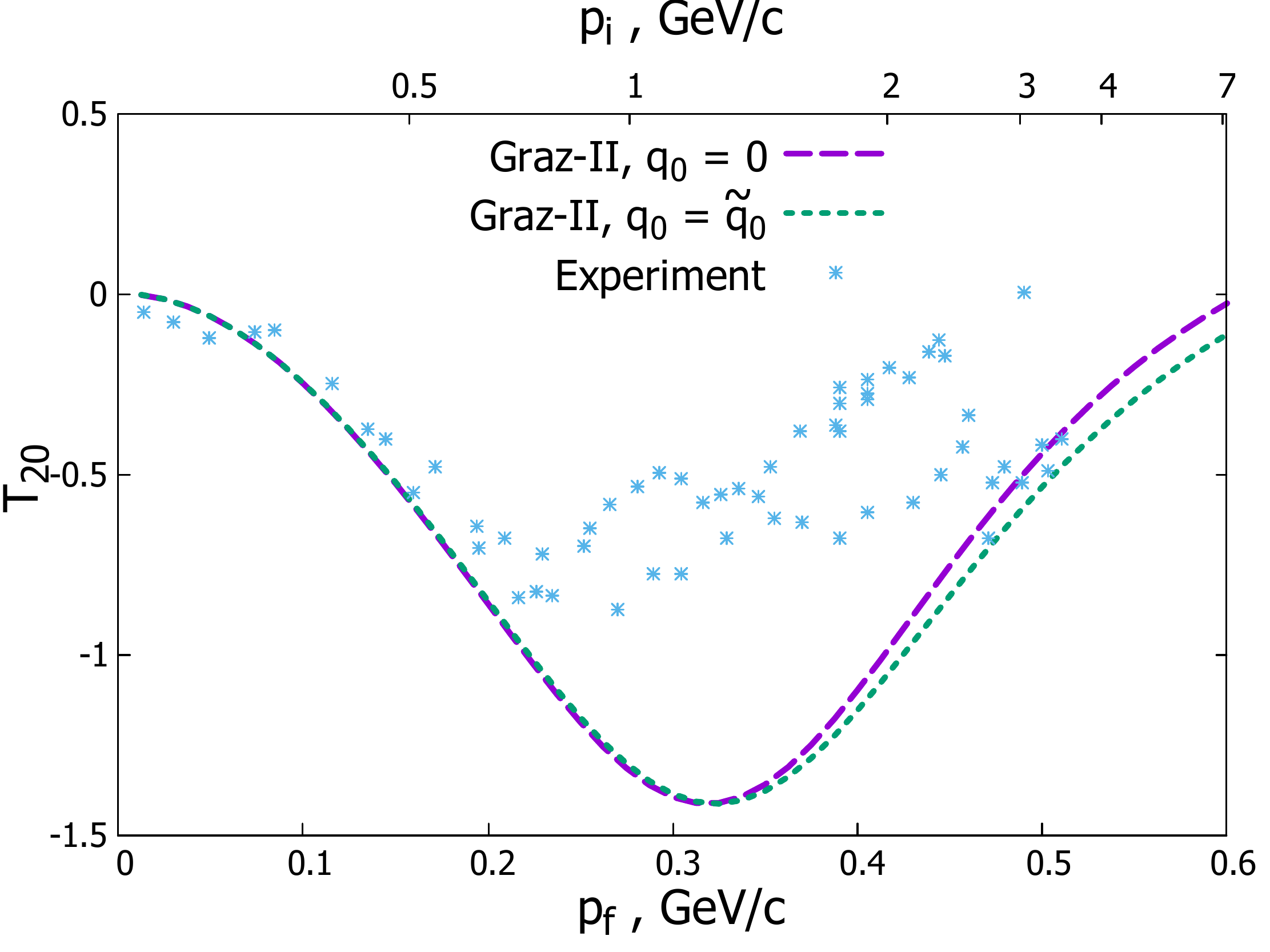}&
	\includegraphics[width=0.45\linewidth,angle=0]{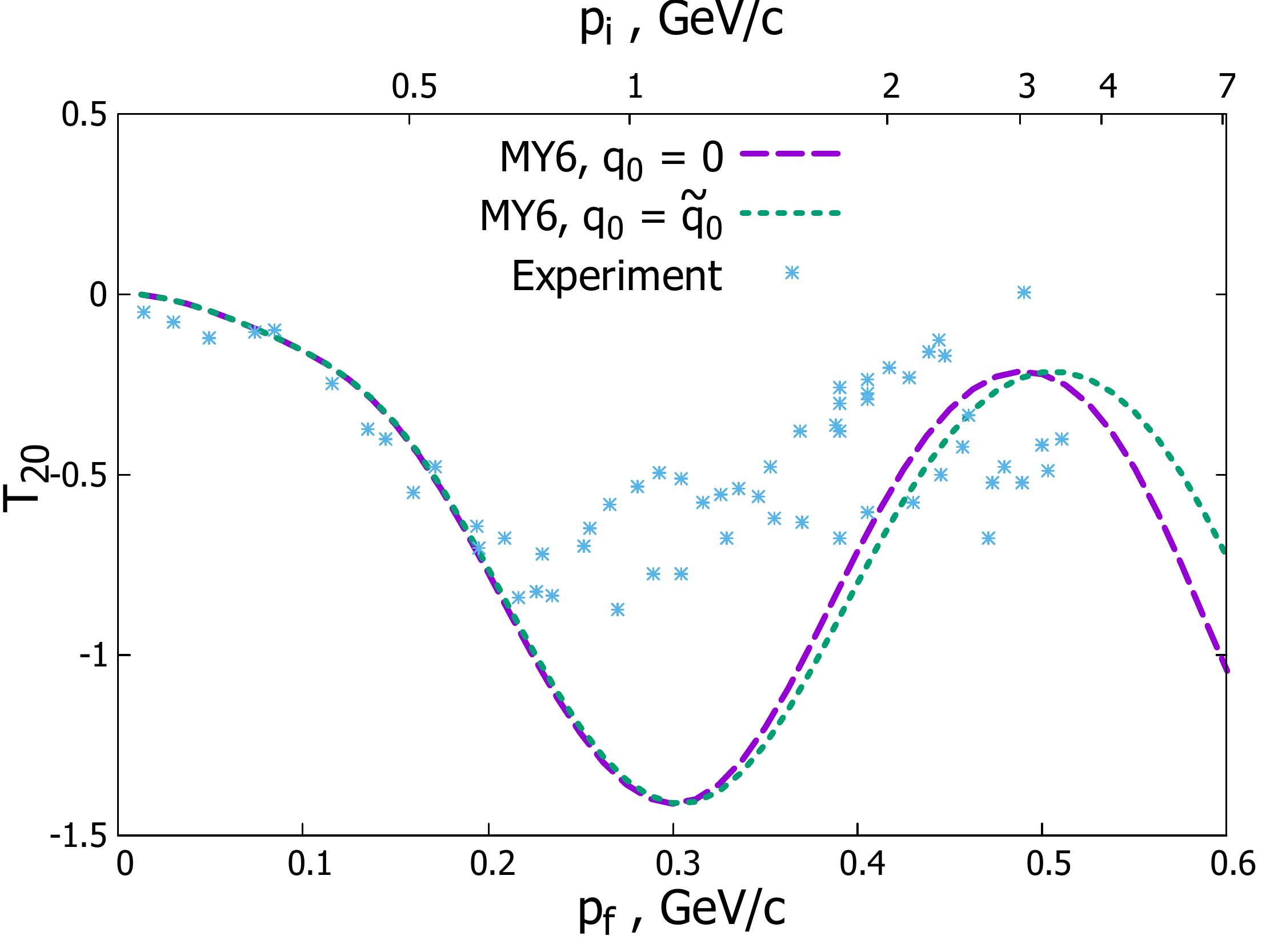}\\
	\includegraphics[width=0.45\linewidth,angle=0]{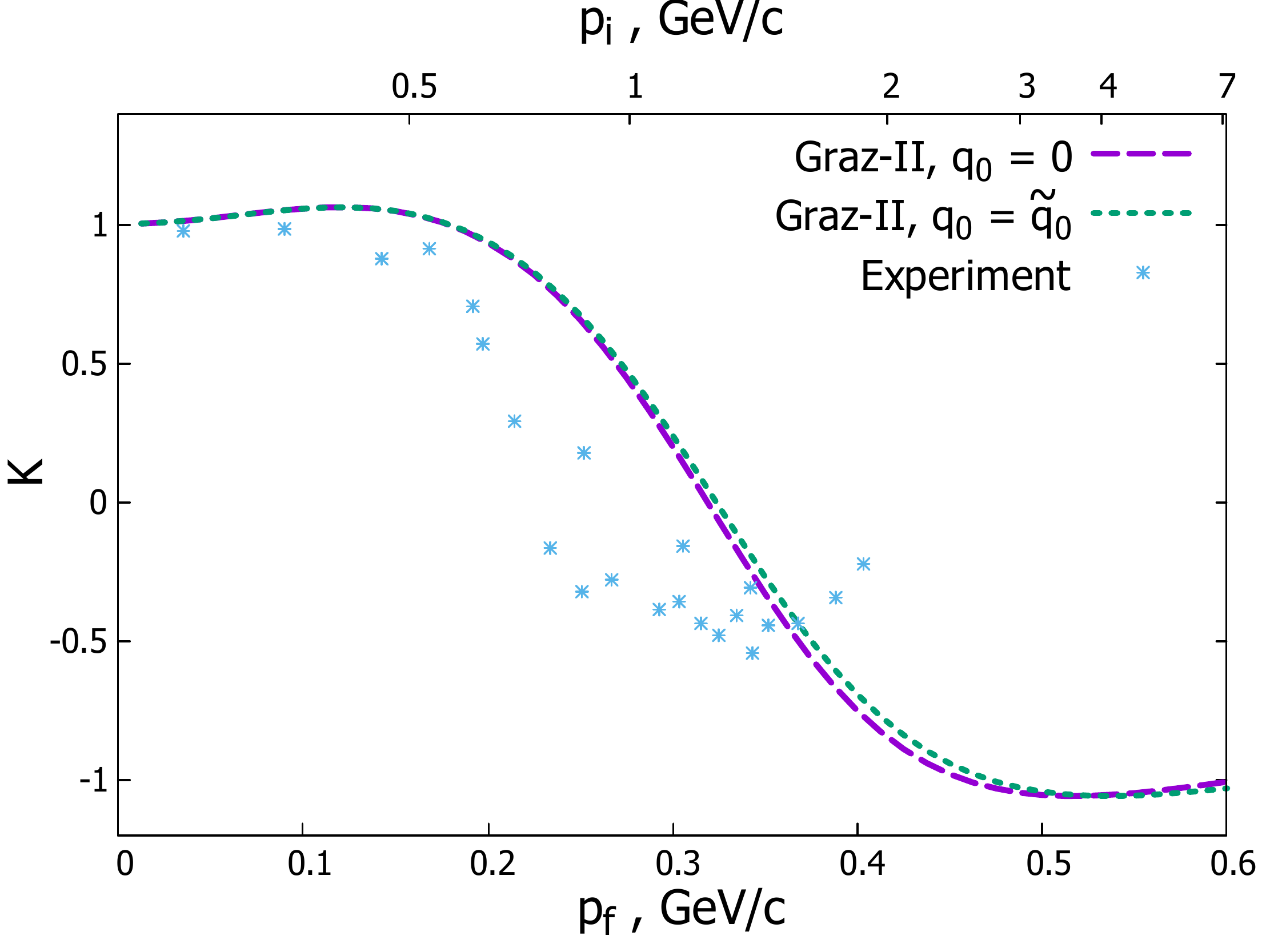}&
	\includegraphics[width=0.45\linewidth,angle=0]{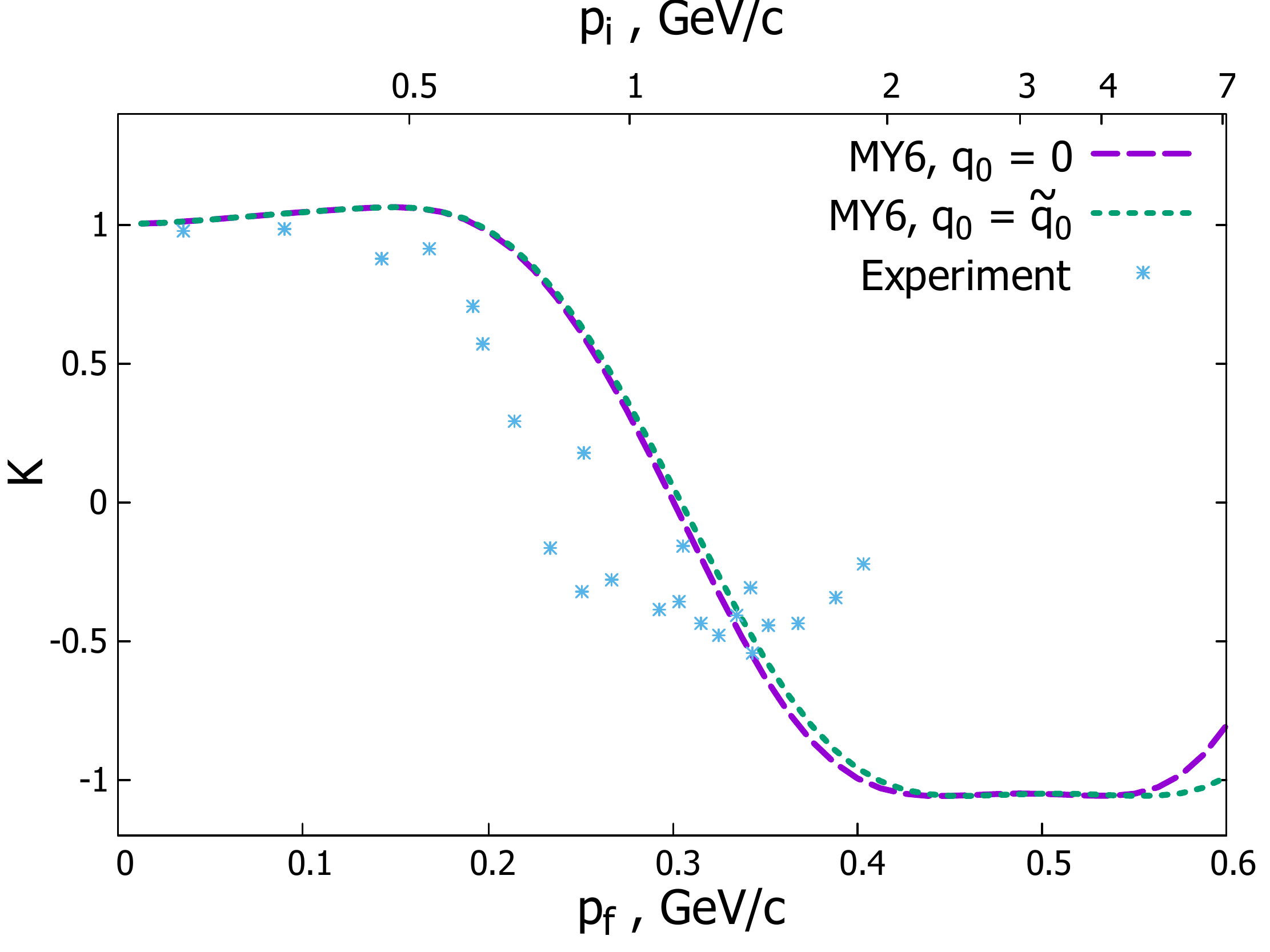}	
\end{tabular}
}
\caption{Unpolarized differential cross section (upper),
tensor analyzing power $T_{20}$ (middle) and  polarization  transfer  $\kappadp$ (bottom)
calculated with $q_0 = 0$ (dashed line) and $q_0 = {\bar p}_0$ (dotted line) in BS amplitude
for Graz-II (left) and MY6 (right) potentials.
}
\label{fig2}
\end{figure}
\begin{figure}[!htb]
\center{
\begin{tabular}{cc}    
	\includegraphics[width=0.45\linewidth,angle=0]{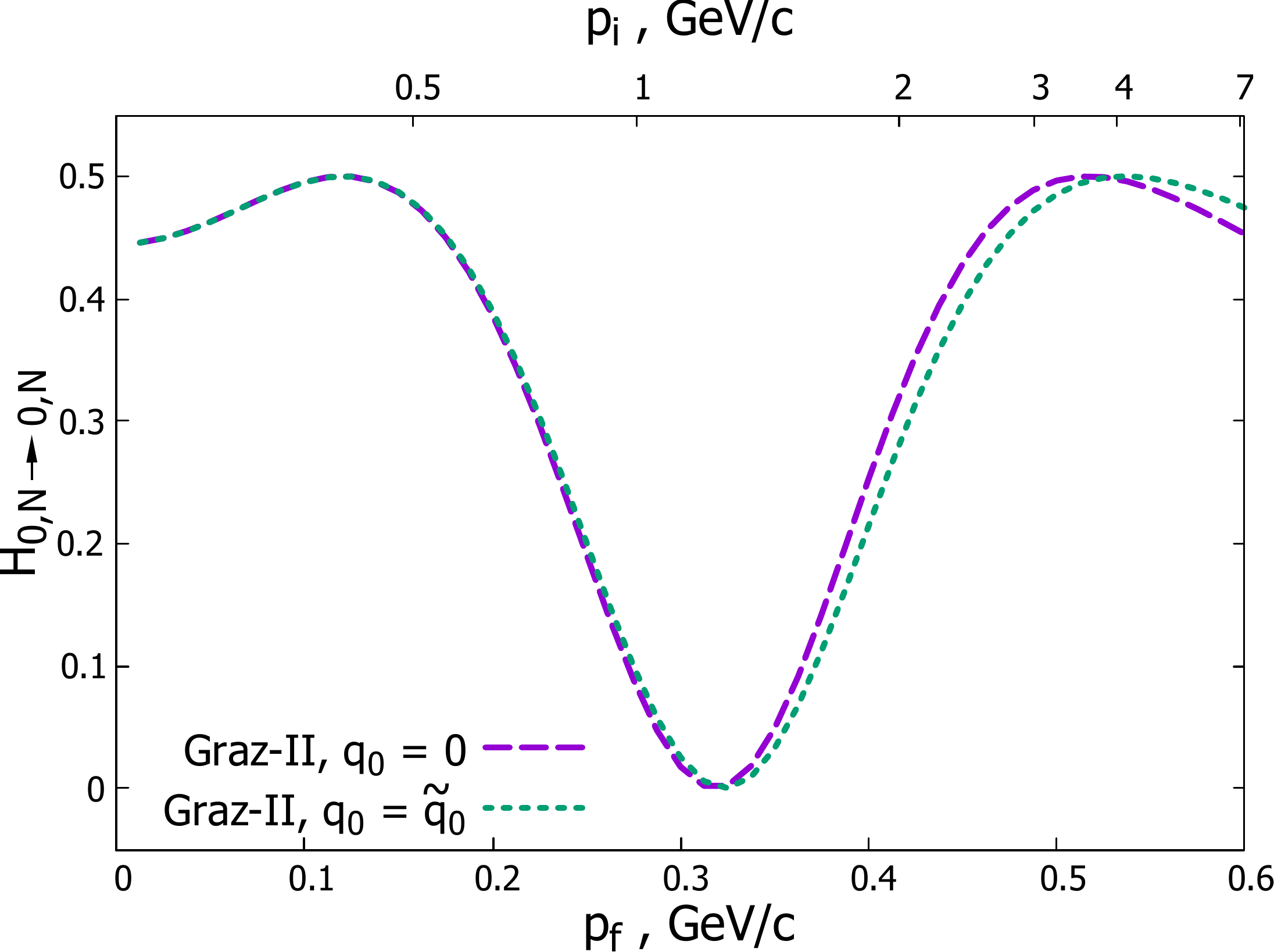}&
	\includegraphics[width=0.45\linewidth,angle=0]{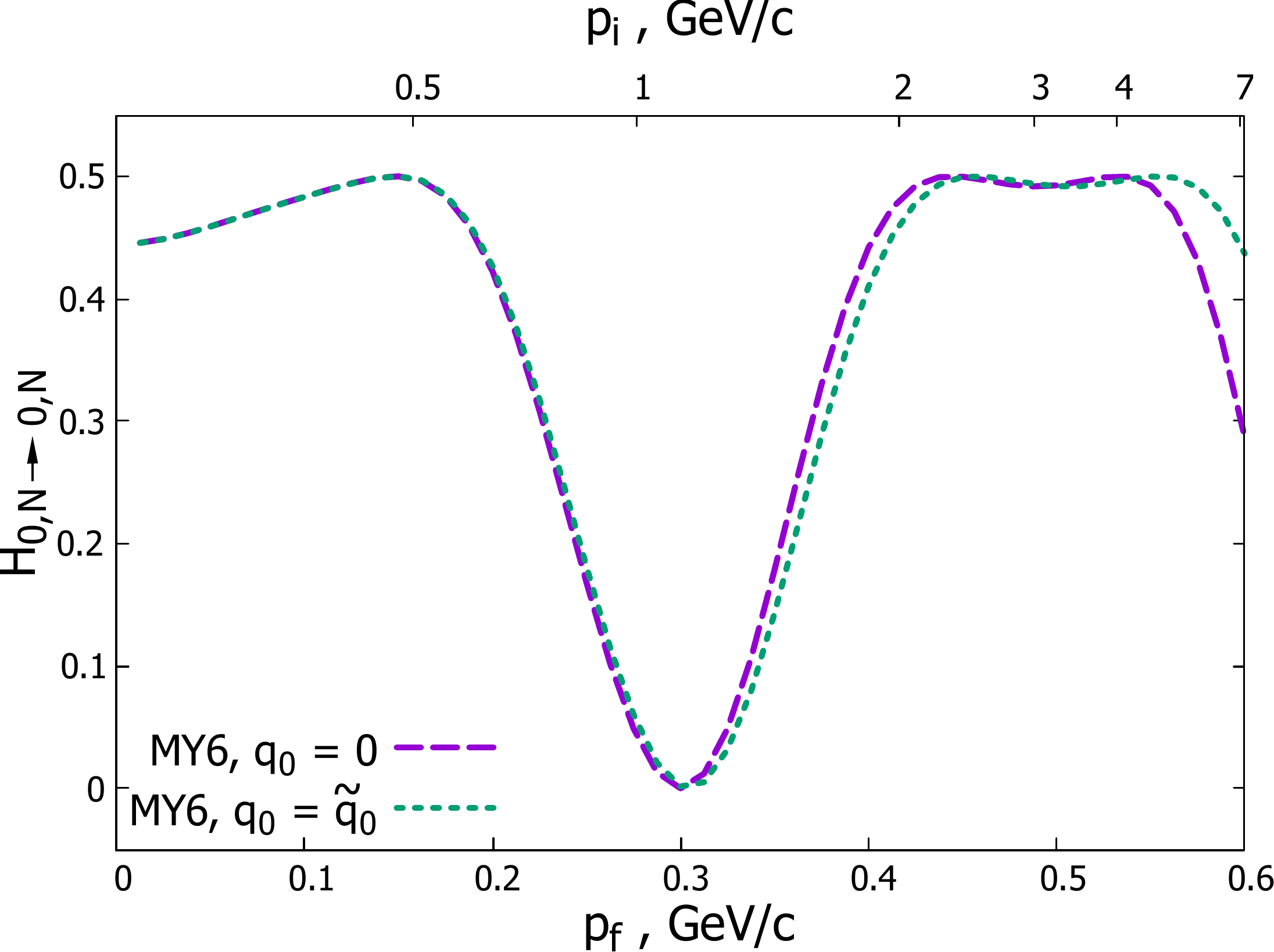}\\
	\includegraphics[width=0.45\linewidth,angle=0]{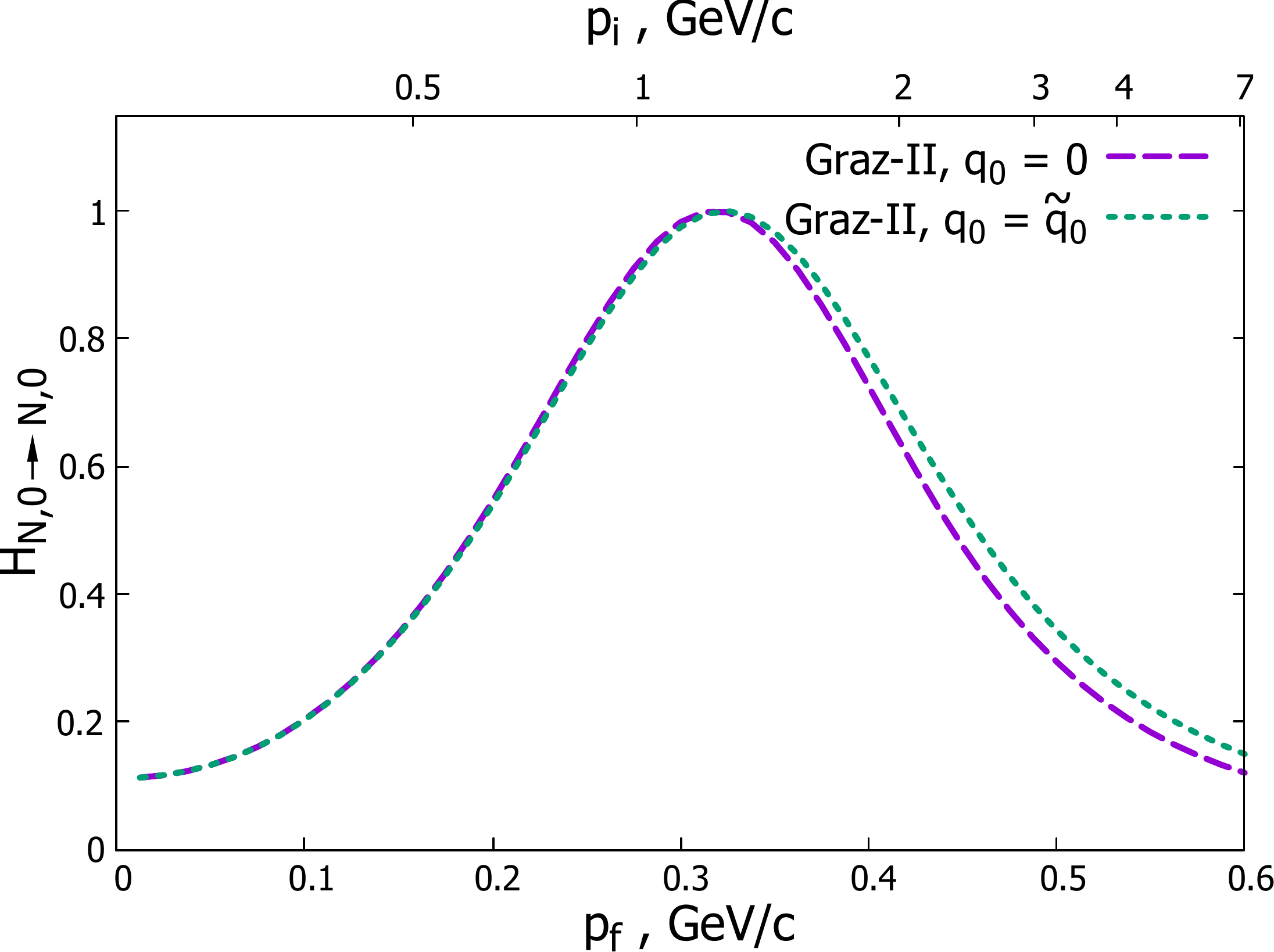}&
	\includegraphics[width=0.45\linewidth,angle=0]{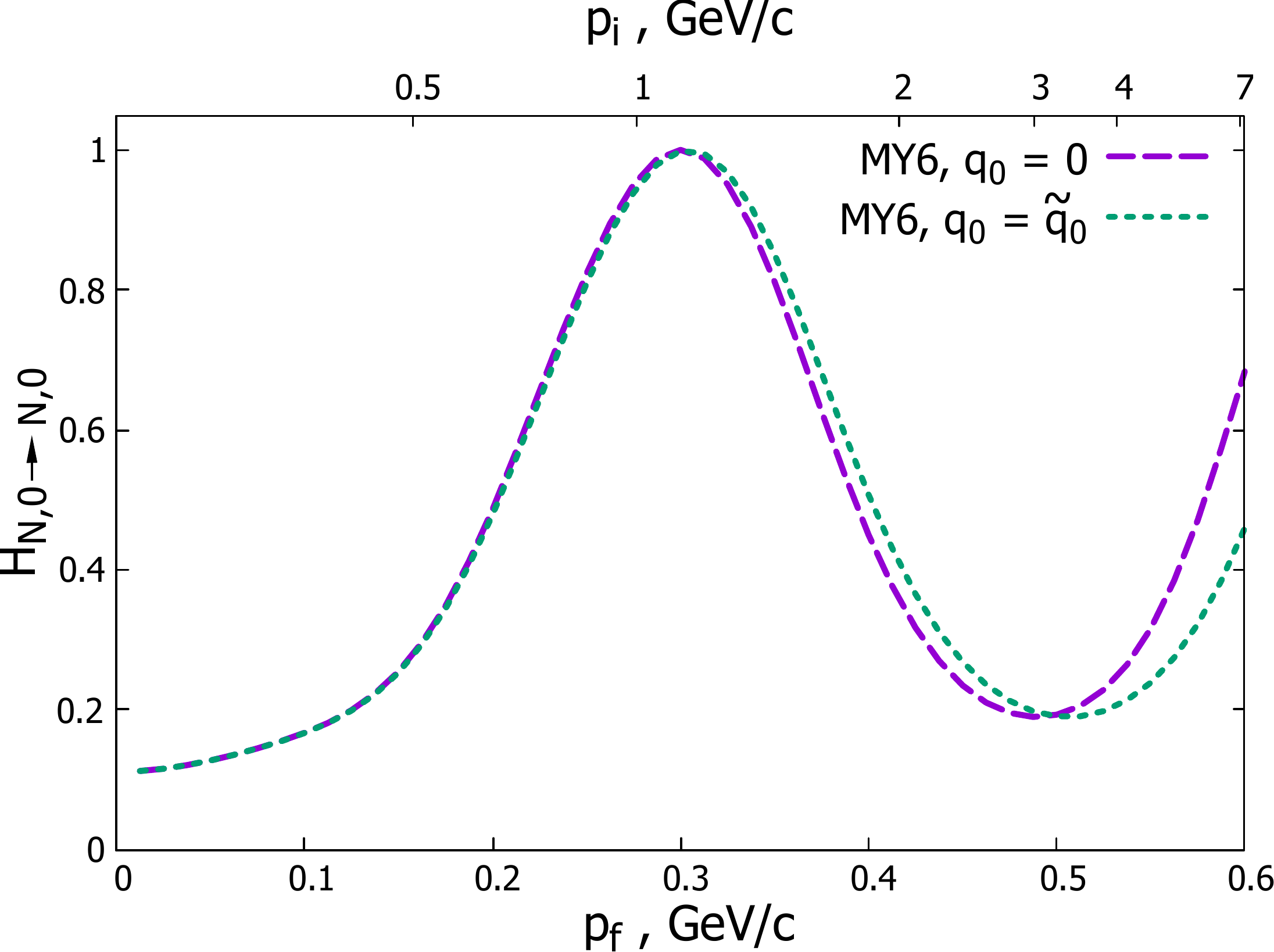}\\
	\includegraphics[width=0.45\linewidth,angle=0]{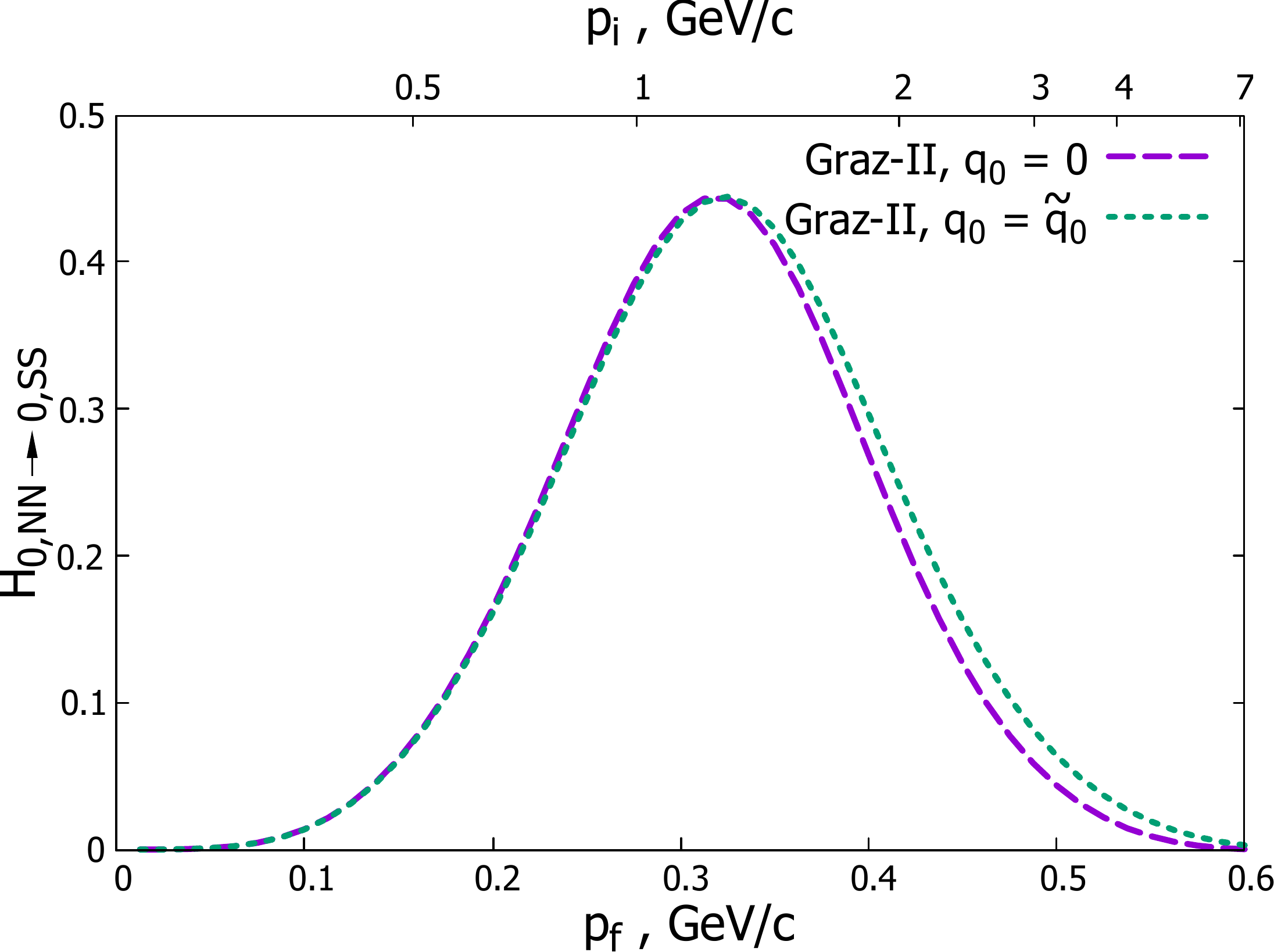}&
	\includegraphics[width=0.45\linewidth,angle=0]{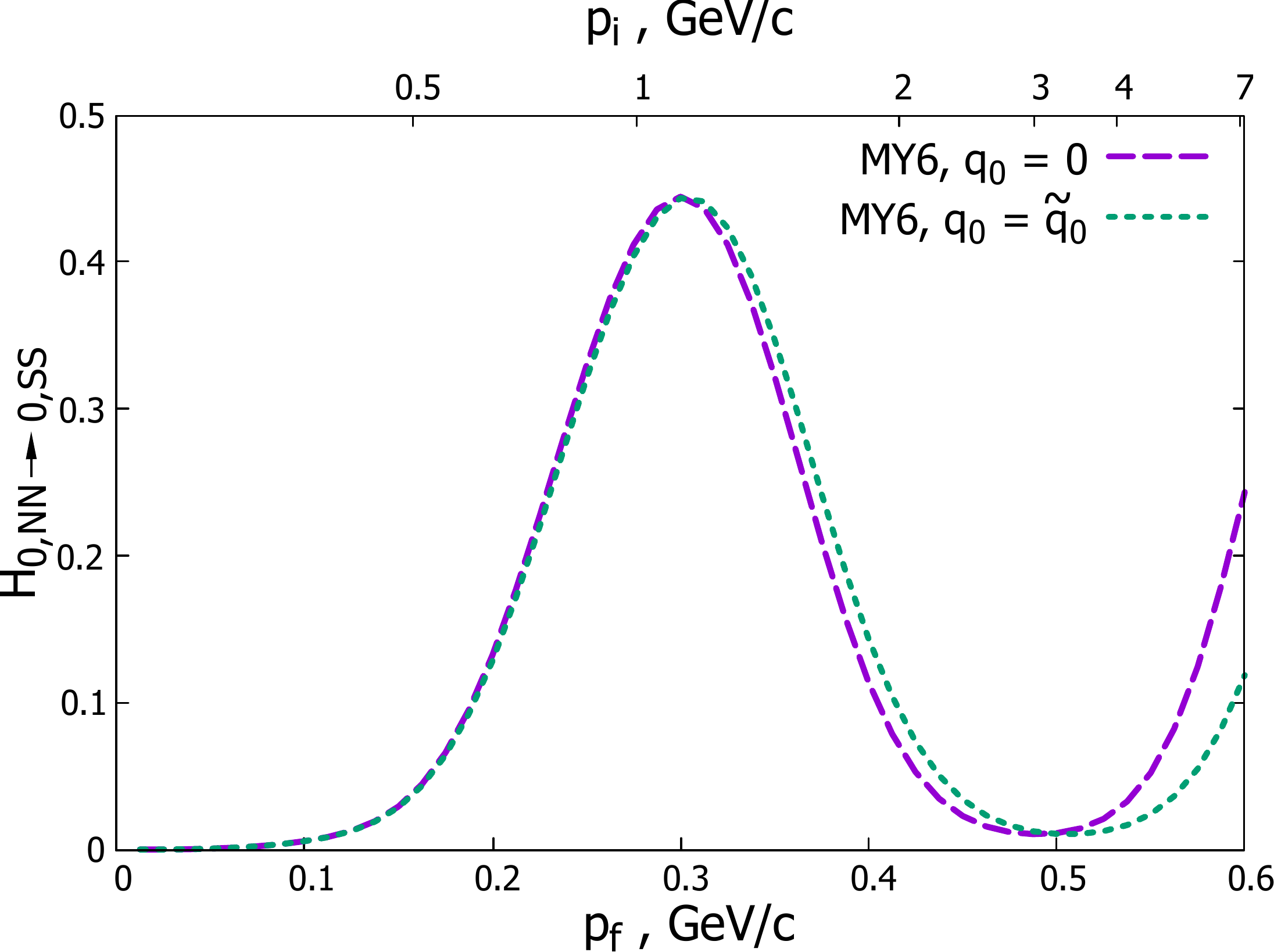}	
\end{tabular}
}
\caption{The same as in Fig.~\ref{fig2} but for $H_{0,N \rightarrow 0,N}$ (upper),
for $H_{N,0 \rightarrow N,0}$ (middle) and for $H_{0,NN \rightarrow 0,SS}$ (bottom).
}
\label{fig3}
\end{figure}

\section{Calculations and results}

Figures~\ref{fig1}-\ref{fig3} show the result of calculations of the differential scattering cross section, tensor and vector polarization characteristics as functions of the final proton momentum $p_{f}$ (up to 0.6 GeV/c) and initial proton momentum $p_{i}$ (up to 7.3 GeV/c). Here and below $|{\bm p_{f}}| \equiv p_{f}$ and $|{\bm p_{i}}| \equiv p_{i}$.

In Figs.~\ref{fig1}, the results obtained in this paper are compared with those from several previous relativistic calculations.
The results from~\cite{KT82} were obtained using the deuteron wave function, that was found by solving the BS equation when one of the nucleons is on the mass shell. The one-meson exchange kernel was considered.

The calculation from~\cite{KD98} was obtained using relativistic wave functions from the solution of the Bethe-Salpeter equation 
with the  one-meson exchange kernel, and then one of the nucleons in the BS amplitude was considered on-mass-shell.
Since both approaches~\cite{KT82,KD98} solve the BS equation in pseudo-Euclidean space, the zero component of the relative 4-momentum 
is set equal to zero.

The relativistic corrections from the Lorentz transformations of the wave function  could also be taken into account 
in the above-mentioned approaches. However, in this paper we omitted them to see the influence of different relativistic kernels 
of the NN interaction.

The results from ~\cite{LA2020} are also presented. They were obtained using the non-relativistic wave function 
of the deuteron with the CD-Bonn interaction potential. 
In this case, Lorentz transformations and Wigner rotations 
were taken into account for the spin part transformation.

The difference between the results of the current paper and results from~\cite{KD98,LA2020,KT82} 
show the influence of the different relativistic potentials as well as different applied approaches.
In these figures, the zero component of the relative momentum $q_0=0$ is used for Graz-II and MY6 solutions. In the calculations, the version of the Graz-II kernel with $p_D=5\%$ is used since all other models have  $p_D$ close to 5\%.

The first picture in Fig. 1 shows the unpolarized differential scattering cross section. One can see that the results are close to each other up to $p_f = 0.30-0.33$ GeV/c. At $p_f =$0.6 GeV/c the result from~\cite{KD98} and the calculation of this paper differ by 3 times. 
For $T_{20}$ a significant difference in  calculations starts approximately at
$p_f =$0.3 GeV/c and  even has an opposite sign at $p_f =$0.6 GeV/c. 
Three of the approaches practically coincide in the region $p_f = 0.15-0.45$ GeV/c,
whith could  probably be explained by using  one-meson exchange type of the NN interaction kennel.
For $\kappadp$, the situation is approximately the same but
the difference with the Graz-II and MY6 calculations starts even at $p_f = 0.20-0.25$ GeV/c.

For the VVPT and TTPT coefficients, only the calculations of~\cite{KD98} and this paper are shown. It is seen that the large difference is observed after $p_f =$0.3 GeV/c where
the VVPT and TTPT coefficients have maximum (minimum) values.

It should be stressed that for the unpolarized cross section, $T_{20}$ and $\kappadp$ observables the description of the experimental data is very poor. It could be considered as an indication of the inadequacy of the considered one-nucleon exchange mechanism of the reaction.

In Figs.~\ref{fig2}-\ref{fig3}, the dependence on the zeroth component of the relative momentum in the BS amplitude is studied. In the figures, the left (right) column stands for the calculations of the observables for Graz-II (left) and MY6 (right) with $q_0 = 0$ and $q_0 = {\tilde q}_0.$ 

The difference in the $q_0 = 0$ and $q_0 = {\tilde q}_0$ cases is negligible for the 
unpolarized cross section but becomes sizeable for the polarization characteristics
especially with increasing  $p_f$ momentum.

It is seen from Fig.~\ref{fig3} that for $T_{20}$ there is no significant difference in the results up to $p_f =$0.3 GeV/c,  but then the difference reaches 0.1 for the potential Graz-II and 0.3 for potential MY6 at $p_f =$0.6 GeV/c.
For $\kappadp$ the difference starts at $p_f =$0.25 GeV/c but the magnitude is smaller than that for $T_{20}$.

For the VVPT and TTPT coefficients in Fig.~\ref{fig3}, the difference between $q_0 = 0$ and at $q_0 = {\tilde q}_0$ starts at $p_f \approx $0.3 GeV/c.

It should be stressed that for several observables taking into account effect of $q_0 \neq 0$ is compatible or even larger than the Lorentz transformation given in~\cite{KD98}
and both of them should be considered simultaneously.

\section{Summary}
	
In this paper, the elastic proton-deuteron backward scattering was considered. The study of this process was carried out within the framework of the relativistic approach based on the one-nucleon exchange.
The unpolarized scattering cross section and polarization characteristics of the process were found for the momentum of the final proton up to 0.6 GeV (up to 2.4 GeV for the momentum in the center-of-mass system and up to 7.3 GeV for the momentum of the initial proton).
The relativistic separable potentials Graz-II and MY6 were used for calculations and the results were compared with other approaches.
The influence of the nonzero component of the 4-momentum in the BS amplitude was also considered. For some polarization characteristics, this effect is sizeable and comparable with the Lorentz boost BS amplitude transformation.

Further improvements are possible by taking into account, in addition to the one-nucleon exchange diagram, 
other diagrams with single and double scattering, taking into account the exchange of $\pi$ mesons and $\Delta$ isobars~\cite{LA2020, KD98}, etc. The most complete account of all diagrams is possible when solving the relativistic Faddeev equation. One can also expect the influence on the observed values of taking into account $P$
partial-wave states in addition to the $S$ and $D$ states considered in this paper.
	

\end{document}